\documentclass[fleqn,twoside]{article}
\usepackage{espcrc2}
\usepackage{graphicx}

\newcommand{\AmS}{{\protect\the\textfont2
  A\kern-.1667em\lower.5ex\hbox{M}\kern-.125emS}}

\newcommand{\autcolor}[1]{{ #1}}
\title{The Static Potential to $O(\alpha^2)$ in Lattice Perturbation Theory
}

\author{
P.\ Boyle\address[ED]{
Department of Physics and Astronomy, The University of Edinburgh, EH9 3JZ, UK
}
\address[COL]{Department of Physics and Astronomy, Columbia University, New York, USA}
and G.S.\ Bali\address[GLA]{Department of Physics and Astronomy,
University
of Glasgow, G12 8QQ, UK}
}
       
\begin{document}
\begin{abstract}
We present a calculation of Wilson loops, and the static
inter-quark potential to $O(\alpha^2)$ in lattice perturbation
theory. This is carried out with the Wilson, Symanzik-Weisz, and
Iwasaki gauge actions and the Wilson, Sheikholeslami-Wohlert, 
and Kogut-Susskind dynamical fermion action for small Wilson loops, and with
the Wilson gauge action and each of the dynamical quark actions in the
case of the static potential. 

\vspace{-.2in}
\end{abstract}
\maketitle

\section{INTRODUCTION}
\vspace{-.05in}
The calculation of Wilson loops and derived quantities (such as
the potential and static quark mass renormalisation) 
is perhaps one of the most mature 
problems in lattice perturbation theory,
with many calculations throughout the last two decades. 

Nevertheless, as the lattice community moves to non-perturbative
simulations with (massive) improved dynamical quark actions it is 
necessary to extend the coverage of the literature correspondingly.

Further, while the calculation of the static  potential has been 
well understood in the literature 
\cite{HellerKarschPotl,WeiszWohlertPotl,WeiszWohlertPotl1}, and
in particular for quenched SU(2) gauge theory \cite{AltevogtGutbrot}, 
we present results for the static potential in $SU(3)$ both with and
without dynamical fermions.
\vspace{-.15in}

\section{Perturbative Wilson Loops}
\vspace{-.05in}
The bare operator for the Wilson loop contains all 
orders in the coupling due to the exponentiation of the gauge
field in a link. This operator must be expanded and 
truncated to $O(\alpha^2)$.
\begin{equation}
W({\cal C}) = \langle {\cal O}_W \rangle 
\end{equation}
\begin{equation}
{\cal O}_W({\cal C}) = \frac{1}{N} {\rm Tr} \prod\limits_{l\in{\cal C}} U_{\mu_l}(x_l) = \prod\limits_l e^{iagA^b_{\mu_l}(x_l)T^b}
\end{equation}
To $O(\alpha^2)$ there are non-trivial contributions from
two, three and four insertions of the gluon field operator at all links
in the perimeter of the loop. 

%
%
%
%
%

We can derive both the static inter-quark potential and the static
quark mass renormalisation from the Wilson loop,
\begin{eqnarray}
V(\vec{R},am) &=& \lim\limits_{T\to\infty} - \frac{\partial}{\partial T} \log W(\vec{R},T)\\
& =& V_{\rm phys}(\vec{R},am) + V_{\rm self}(am) \nonumber
\end{eqnarray}
\vspace{-.3in}
\begin{equation}
V_{\rm self} = \lim\limits_{|\vec{R}|\to\infty} V(\vec{R})
\end{equation}

The temporal component of the internal loop momentum 
was analytically integrated out to pick out the pole structure,
and then the remaining integrals were performed in 
the infinite volume limit using VEGAS. In the case of the potential and 
the $V_{\rm self}$ the limits in coordinate space produce additional 
delta functions in momentum space which were integrated over analytically.
\vspace{-.1in}
\section{Small Wilson Loops}
\vspace{-.05in}
For pure gauge, we write the Wilson loop as,
\begin{eqnarray}
W_{PG} &=& 1 - g^2 \frac{N^2-1}{N} W_T \\
&&- g^4 [ (N^2-1) W_1 + \frac{1}{ N^2 } W_2 ] \nonumber
\end{eqnarray}
In the dynamical case we write,
\begin{equation}
\begin{array}{ll}
W &= W_{PG} 
- g^4 n_f  \frac{N^2-1}{N} \\&\times\left[ X_f^{(0)}+X_f^{(1)} c_{sw} +
X_f^{(2)} c_{sw}^2 \right] 
\end{array}
\end{equation}
for Wilson type fermions and,
\begin{equation}
 W  = W_{PG} - g^4 \frac{N^2-1}{N} n_f X_f^{(KS)} 
\end{equation}
for Kogut-Susskind fermions.

We obtain new results for non-planar Wilson loops in the pure gauge case
in Table~\ref{TabSmallPG}, and for various dynamical fermion actions
in Table~\ref{TabSmallFermions}. The results for planar small Wilson loops
in pure gauge theory are well known\cite{HellerKarschPotl,WeiszWohlertPotl,PanaFeo,IsoSakai}.

\begin{figure}[htb]
\vspace{-.2in}
\caption{
\label{FigMassPlaquette}
Mass dependence of the fermionic contributions to the plaquette
for Wilson, Clover and Kogut-Susskind fermions. The stand-alone
points correspond to the $am=0$ limit.
}
\includegraphics[width=7cm]{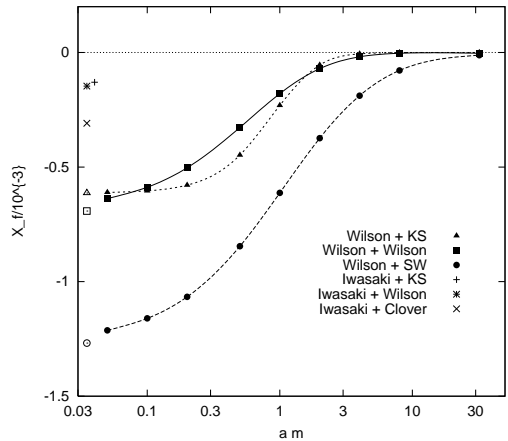}
\vspace{-.2in}
\end{figure}

\begin{table}[hbt]
\vspace{-.25in}
\caption{
\label{TabSmallPG}
Pure Gauge Small Wilson Loop Results with Wilson Gauge Action}
\begin{tabular}{cccc}
\hline
Loop  & $W_T$ &  $W_1$ & $W_2$ \\
\hline
\includegraphics[width=0.5cm]{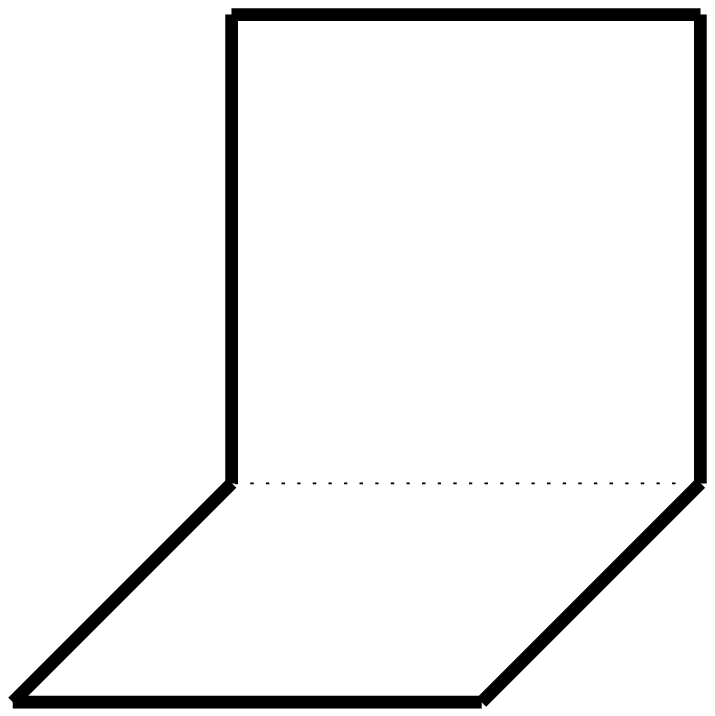}&
    \autcolor{0.1961(1)}&\autcolor{0.002624(1)}&\autcolor{0.005284(5)}\\
\includegraphics[width=0.5cm]{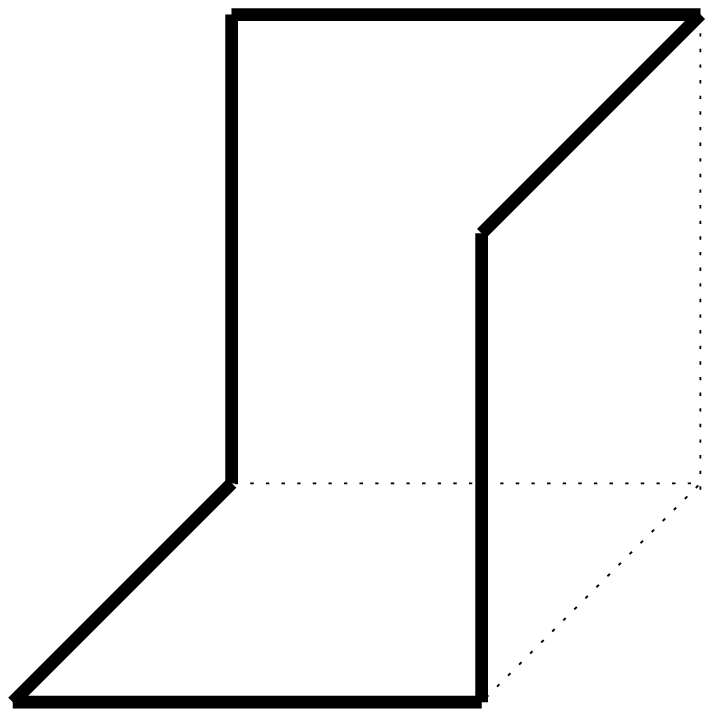} & 
\autcolor{0.2134(1)} & \autcolor{0.001845(1)} & \autcolor{0.003911(5)} \\
\hline
\end{tabular} 
\vspace{-.3in}
\end{table} 
\begin{table*}[hbt]
\caption{
\label{TabSmallFermions}
Fermionic contribution to loops (various Quark \& Gluon actions)}
\begin{tabular}{cccccc}
\hline
Glue & Loop &$ X_{nf}^{(0)} (am=0)$ & $X_{nf}^{(1)} (am=0)$ & $X_{nf}^{(2)} (am=0)$ & $X_{nf}^{(KS)}(am=0)$\\
\hline
Wilson & $1\times 1$    & $-6.96(2)\times 10^{-4}$  
           & \autcolor{$2.02(2)\times 10^{-5}$}   
           & \autcolor{$-5.963(1)\times 10^{-4}$} 
           & $-6.129(3)\times 10^{-4}$ \\
Iwasaki & $1\times 1$ &  \autcolor{$-1.47(1)\times 10^{-4}$}
           & \autcolor{$ 7.8(1)\times 10^{-7}$}
           & \autcolor{$-1.698(1)\times 10^{-4}$} 
	   & \autcolor{$-1.296(2)\times 10^{-4}$} \\
Wilson &\includegraphics[width=0.5cm]{chair.eps}
& \autcolor{$-1.487(4)\times 10^{-3}$} & \autcolor{$3.62(5)\times 10^{-5}$} 
& \autcolor{$9.836(3)\times 10^{-4}$} & \autcolor{$1.337(1)\times 10^{-3}$}\\
Wilson & \includegraphics[width=0.5cm]{parallelogram.eps}
& \autcolor{$-1.653(6)\times 10^{-3}$} & \autcolor{$5.04(6)\times 10^{-5}$} & 
  \autcolor{$1.1624(3)\times 10^{-3}$} & \autcolor{$1.445(1)\times 10^{-3}$} \\
Wilson & $1\times 2$				     
	& $-1.326(3)\times 10^{-3}$ & \autcolor{$5.67(5)\times 10^{-5}$} & \autcolor{$1.4759(3)\times 10^{-3}$} & $1.179(1)\times 10^{-3}$\\
Wilson & $2\times 2$				     
	& $-2.38(1)\times 10^{-3}$  & \autcolor{$1.61(1) \times 10^{-4}$} & \autcolor{$3.3421(6)\times 10^{-3}$} & $2.056(2)\times 10^{-5}$\\
\hline
\end{tabular}
\vspace{-.2in}
\end{table*}
The extraction of $\alpha_s$ with massive dynamical quarks
in a ``partially quenched'' analysis, where the scale is set before
extrapolation in the quark mass, is somewhat subtle since it
depends on both the perturbative plaquette, and on
the connection to $\overline{MS}$.
In Figure~\ref{FigMassPlaquette} we show the mass 
dependence of the perturbative plaquette 
with the Wilson gauge action and both (improved) Wilson quark 
and Kogut-Susskind quark actions with massive dynamical quarks. 
This shows a smooth connection to the quenched case as the 
mass is increased across the threshold $ma \simeq 1$.
These results combined with the (lattice) mass dependent connection
between the lattice and $\overline{MS}$ schemes discussed in \cite{inprep}
allow a calculation of $\alpha_s(m_q,\mu)$.

\section{Potential}

We define the lattice coordinate space potential coupling
\begin{equation}
\alpha_R(a\vec{R},am) = - \frac{aR}{C_F} \left( V(a\vec{R},am) - V_{\rm self}(am) \right)
\end{equation}
\begin{equation}
\begin{array}{cc}
V(a\vec{R},am)=& \alpha_L V_1(a\vec{R}) + \alpha^2_L \\ \times &\left( V_2^{PG}(a\vec{R}) + n_f V_2^F(a\vec{R},am)\right) 
\end{array}
\end{equation}

\begin{figure}[htb]
\vspace{-.7in}
\caption{
\label{FigQuenchedPotl}
The one-loop potential. The lattice tadpole
contribution is exactly proportional to the tree level potential.
The one loop terms run with the appropriate part of the logarithmic beta function.
}
\includegraphics[width=7cm]{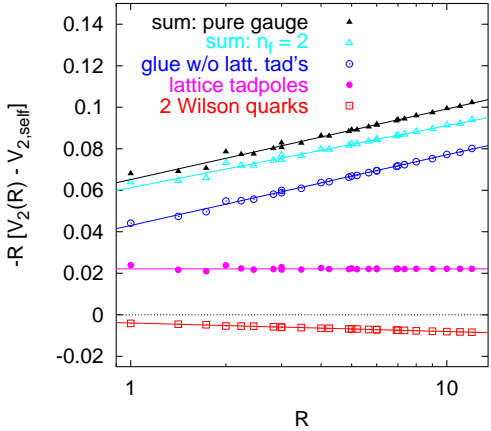}
\vspace{-.4in}
\end{figure}

$\alpha_R(a\vec{R},am)$ is constant up to discretisation errors at tree level.
At one loop it runs logarithmically with $R$, with  pure gauge
and fermionic contributions.
In pure gauge we can identify a lattice tadpole
at one loop, shown in Figure~\ref{FigQuenchedPotl}, 
which is exactly proportional to the tree level contribution.
The improvement of lattice potentials through the use of
a ``lattice-R'' arises at this order through the obvious correlation of 
the one loop contribution with the tree level one.

\begin{figure}[htb]
\vspace{-.35in}
\caption{
\label{FigFermAll}
Fermionic contributions to the coordinate space coupling
with the Wilson gauge action and the Clover, Wilson, and 
Kogut Susskind actions. Decoupling is observed in the massive case.
$n_f$ is two for Wilson and Clover and four for Kogut-Susskind.
}
\includegraphics[width=7cm]{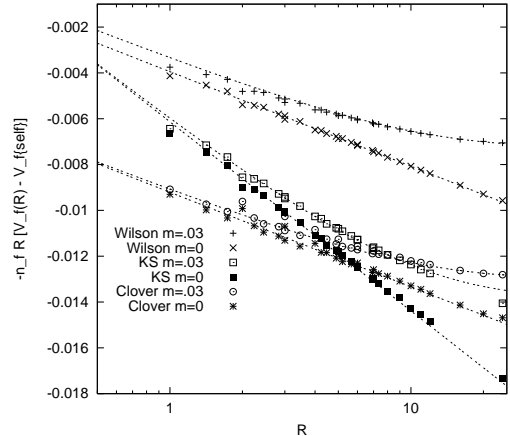}
\vspace{-.35in}
\end{figure}

Figure~\ref{FigFermAll} 
shows some sample fermionic contributions to the coordinate space
potential for each of the quark actions studied both in the massless
limit and with a non-zero quark mass. The smooth curves are continuum 
predictions\cite{Melles}, with the known conversion from $\overline{MS}$ 
used in the massless case. 
In the massive case we obtain the conversion
from $\overline{MS}$ to this scheme from a fit to the data \cite{inprep}.
The R-scheme is a physical one and the decoupling of the quarks from the
running at $aR \gg (am)^{-1}$ can be seen.

We find that the BLM $q^\ast$ for the potential \emph{including} the self
energy term is well described by
\begin{equation}
aq^\ast(aR) =  1.701 + \frac{1.1479}{aR} + \frac{0.20333  \log aR}{a R}
\end{equation}
If we seek to predict the short distance potential we should look
either at the force or a difference in potentials such that the 
the unphysical and ultraviolet self energy terms cancel.
The constant term in the above $q^\ast$ cancels and the scale runs
rapidly with $R$. 
In such combinations the perturbation theory performs remarkably well
compared to non-perturbative data for the short and intermediate distance
potential \cite{inprep,Necco:2001gh}. Also, this work obtains an understanding
of discretisation effects on the short distance potential.
This is essential in order to discern dynamical quark effects 
on the short distance potential.

\section*{ACKNOWLEDGEMENTS}
\vspace{-.1in}

P.B.\ is supported by PPARC grant PPA/J/S/1998/00756.
G.B.\ is a Heisenberg Fellow (DFG grant Ba~1564/4-1).            
This work is supported by PPARC grant PPA/G/O/1998/00559 and
by the EU networks HPRN-CT-2000-00145 and FMRX-CT97-0122.

\end{document}